# Flux enhancement in the inner region of a geometrically and optically thick accretion disk

R. Misra[1] and K. Sriram[2]

## ABSTRACT

The surface flux (and the corresponding observed flux) of a geometrically thick "funnel" shaped disk is computed taking into account the radiation impinging on the surface from other parts of the disk. It is found that the ratio of the maximum apparent luminosity to the real luminosity of the disk is only a factor $\approx 5$ even when the opening angle of the disk is small ($\approx 15^o$). Thus, geometrically beamed emission from "funnel" shaped sub-Eddington disks around stellar mass black holes, cannot explain the Ultra-Luminous X-ray sources detected in nearby galaxies.

*Subject headings:* accretion, accretion disks—black hole physics

## 1. Introduction

Recent observations by the Chandra Observatory of nearby galaxies have confirmed the presence of Ultra-Luminous X-ray objects (ULX) ( e.g. Kaaret *et al.* 2001; Matsumoto *et al.* 2001; Zezas & Fabbiano 2002 ) which were detected earlier by ROSAT (e.g. Roberts & Warwick 2000). A large number of ULX, that are defined as off center X-ray sources with luminosities $> 10^{39}$ ergs/sec, were detected by ROSAT (Colbert & Ptak 2002). Since the luminosities of these sources are greater than the Eddington luminosity ($L_{Edd} \equiv 4\pi GM m_p c/\sigma_T$) for a ten solar mass black hole, it has been proposed that they harbor intermediate mass black holes (IMBH) with a mass range of $10^2 - 10^5 M_\odot$. The upper limit on the black hole mass is set by dynamic friction which would drive a larger mass black hole to the center of the galaxy in a Hubble time (Kaaret *et al.* 2001). Some ULX are variable and show soft to hard state transitions similar to stellar mass galactic black hole systems (Kubota *et al.* 2001). Colbert & Mushotzky (1999) found that the ASCA spectral data of some of the ULX discovered by ROSAT can be well fit by the bulk motion Comptonization model (e.g. Shrader & Titarchuk 1998). They used the spectral fit parameters to directly estimate the mass of the black holes in these systems using the method developed by Borozdin *et al.* (1999) and obtained masses in the range of $10^2 - 10^4 M\odot$, which indicated that these systems harbor IMBH.

[1]Inter-University Center for Astronomy and Astrophysics, Post Bag 4, Ganeshkhind, Pune-411007, India; rmisra@iucaa.ernet.in

[2]Dept. of Astronomy, Osmania University, Hyderabad-500007, India

The existence of IMBH leads to a paradigm shift from the earlier view that there are two kinds of black holes ( stellar and super-massive) in the universe. The formation mechanism of IMBH is uncertain and several possible mechanisms have been suggested. Portegies Zwart & McMillian (2002) have proposed that stellar collisions in a dense cluster could lead to the formation of IMBH. An IMBH may also have been formed by the merger of neutron stars or smaller black holes that are the remnants of massive stars in molecular clouds (Taniguchi *et al.* 2000). On the other hand, Madau & Rees (2001) have suggested that IMBH are remnants of massive ( $> 300 M_\odot$) primordial stars. These models have to be confirmed by detailed simulations and their predictions regarding number density and location of ULX have to be compared with observations ( e.g. Zezas & Fabbiano 2002). Apart from the formation issues, there are also uncertainties as to how these IMBH occur in binaries. King *et al.* (2001) argue that since standard co-evolution may not be possible, these IMBH have to form binaries by stellar capture which should be a rare event. They further argue that in such systems it would be difficult to maintain persistent levels of high accretion. Considering these uncertainties it is prudent to consider alternate models for ULX.

One alternate model for ULX, is that they are stellar size black hole systems with super-Eddington luminosities. It has been difficult to construct steady state models for super-Eddington luminosities. Recently, Begelman (2002), have proposed that a disk may have super-Eddington luminosities, if there are time-varying inhomogeneities due to the photon-bubble instability. This interesting result also needs to be confirmed both observationally and theoretically by time-dependent numerical simulations. A simpler alternate model is that ULX are sub-Eddington stellar mass black hole systems with beamed emission. Since some of the spectra of ULX have a dominant thermal component, the emission cannot arise from a relativistic jet. However, as pointed out by King *et al.* (2001), there could be significant beaming if the emission is from the inner regions of a geometrically thick disk. If the geometry of the disk confines the emission to a solid angle $\Delta\Omega$, then the apparent luminosity would be a factor $\chi_L \approx 2\pi/\Delta\Omega$ larger than the real luminosity. If the opening angle of the disk is $\alpha$ then $\chi_L \approx 1/(1 - \cos\alpha)$ and hence $\chi_L > 10$ for $\alpha < 25^o$. While, such a scenario may not be applicable for the brightest ULX ( $L > 10^{41}$ ergs/sec), it could make the larger number of ULX ( $L \approx 10^{39-40}$ ergs/sec), harbor regular stellar mass black holes ( King *et al.* 2001). Hence the number of IMBH in a galaxy could be significantly smaller then that estimated by taking the entire ULX population into account, which could perhaps alleviate some of the difficulties faced by IMBH formation theories.

In this paper, our motivation is to study and evaluate the observed luminosity enhancement from geometrically thick disks. Our primary goal is to study the feasibility of the hypothesis that ULX are sub-Eddington stellar mass black hole systems with geometrically thick accretion disk and hence we simply assume the structure of the disk instead of solving for it self-consistently. For simplicity we ignore relativistic light-bending effects. In the next section, the assumptions, calculations and results are presented, while the last section is devoted to summarizing the work and discussion.

## 2. Flux from a geometrically thick disk

We approximate the geometry of a thick disk by a bi-cone "funnel" as shown in Figure 1. In this simple geometry the shape of the funnel is specified by the opening angle ($\alpha$), the radius of the marginal stable orbit ($R_{ms}$) and the radial extent of the funnel ($R$). For radii greater than $R$, it is assumed that the height decreases with radius and the flux from this outer region is neglected in the calculation. All through this work $R_{ms}$ is taken to be $6GM/c^2$ (i.e. for a Schwarzschild metric) and $R = 11R_{ms} = 66GM/c^2$. However, the results presented in this work are scale invariant and only depend on the ratio $(R - R_{ms})/R_{ms} = 10$. Thus they would be equally valid for $R_{ms} = 2GM/c^2$ (i.e. for a maximal rotating black hole) and $R = 22GM/c^2$. The value of $R$ has been chosen such that the inner funnel extends to well beyond the region of maximum flux dissipation. Realistic thick disks may not extend to such large radii and hence the luminosity enhancement computed below may be considered as an upper limit. Increasing the value of $R$ does not significantly change the results.

The flux dissipated due to viscosity is assumed to be

$$F_d = \frac{C}{r_h^3}(1 - (\frac{R_{ms}}{r_h})^{1/2}) \qquad (1)$$

where $r_h$ is the horizontal distance from the black hole and $C$ is some arbitrary constant. Although the above expression is strictly valid only for a Keplerian disk, for this work it is adequate since it insures that the maximal flux dissipation occurs deep inside the funnel, which in turn would give the maximum luminosity enhancement. The surface flux at a distance $r = r_h/\sin\alpha$, from the black hole due to this dissipation is

$$F_{sd}(r) = F_d(r_h)\sin\alpha \qquad (2)$$

The total surface flux will be the sum of $F_{sd}$ and the incident flux due to the other parts of

the disk. Hence

$$F(r) = F_{sd}(r) + \int \frac{F(r')}{\pi d^2} \cos\beta \cos\beta' dA' \qquad (3)$$

where $d$ is the distance of the line joining points $r$ and $r'$, $\beta$ ($\beta'$) is the angle between line $d$ and the normal to the surface area $dA$ ($dA'$). Equation (3) is solved iteratively till $F(r)$ converges. The structure of the flow inside the marginal stable orbit will depend on the details of the hydrodynamic flow and accretion rate. Here, we consider two extreme conditions where the flow is optically thin and when it is optically thick. For the optically thin case, it is assumed that radiation from the other side of the disk can go through the region $r < R_{ms}$. Hence the integral in equation (3) includes the contribution from the other side provided both $\cos\beta$ and $\cos\beta'$ are positive. In the other extreme scenario, when the flow is optically thick, it is assumed that there is a non-zero $F(r)$ inside the marginal stable orbit, which is equal to the flux incident on it (i.e. the integral term in eqn. 3). In both cases, we ignore the presence of the black hole. Figure 2, shows $F(r)$ versus the horizontal distance $r_h$ for $\alpha = 15^o$ for both the optically thick and optically thin case. The dissipative flux $F_{sd}$ is also shown for comparison.

The flux observed at earth from such a thick disk is given by

$$F_o(i) = \frac{1}{\pi D^2} \int F(r) \cos\beta_o \mathrm{H}(A, i) dA \qquad (4)$$

where $i$ is the inclination angle of the system with respect to the observer, $D$ is the distance to the source, $\beta_o$ is the angle between the normal to the surface area and the direction to the observer. The Heaviside function $H(A, i)$ is zero if the region ($dA$) is shielded and unity otherwise. The flux enhancement $\chi$ which is the ratio of the apparent luminosity to the real luminosity is

$$\chi(i) = \frac{4\pi D^2 F_o(i)}{L_A} \qquad (5)$$

The real luminosity is given by

$$L_A = 2 \int F_{sd}(r) dA \qquad (6)$$

where the factor 2 takes into account radiation from both sides of the disk. Note that with this definition, $\chi = 2\cos(i)$ for a flat disk. Moreover, by conservation of total energy,

$$\int_0^{\pi/2} \chi(i) d\cos i = 1 \qquad (7)$$

is true for any geometry.

Figure 3, shows the variation of $\chi$ with inclination angle ($i$) for two different values of opening angle $\alpha$. The flat disk value ($\chi = 2\cos(i)$) is also plotted for comparison. $\chi$ is plotted only for the case when the innermost region is optically thin, since the alternate case when the region is optically thick produces a nearly identical curve. This implies that the results presented here are not sensitive to the exact geometry of the innermost regions. To verify the numerical computation, eqn. (7) has been checked for all $\chi$. As discussed in the introduction, for $\alpha = 15^o$, the expected enhancement was $\approx 1/(1 - \cos\alpha) \approx 30$. However, from Figure 3, the computed enhancement is only $\approx 5$ which is nearly an order of magnitude less than the expectation. The reason for this is that the "funnel" disk does not confine the flux to a narrow solid angle, and there is significant flux even when $i > \alpha$ (Figure 3). This flux at large angles is mainly due to the outer regions of the disk which have been illuminated by the inner part.

## 3. Summary and Discussion

In this work, it is shown that a geometrically thick funnel shaped disk does not confine the radiation to a narrow solid angle even when the opening angle is small. There is significant flux at viewing angles greater than the opening angle of the disk which in turn implies that the apparent luminosity of such a disk is only a factor $\approx 2$ higher than that of a flat disk even when the opening angle is as small as $15^o$. Thus sub-Eddington stellar mass ($M < 10 M_\odot$) systems with geometrically thick disks, cannot account for the low luminosity ($L \approx 10^{40}$ ergs/sec) ULX observed by Chandra.

There are two simplifying assumptions made in this calculations which are discussed below. First, radiation is assumed to move in straight lines i.e. light bending effects have been ignored. The effect of light bending could either enhance or decrease the surface flux ($F(r)$) calculated here (eqn. 3) by either bending the radiation toward or away from the surface area. Similarly, the component of the observed flux from the innermost regions would be affected. However, it seems unlikely that the surface flux at the outermost (or the topmost) region of the funnel will be strongly effected by this effect. Hence, there would still be significant flux at inclination angles ($i > \alpha$). Since, eqn (7) is always valid, this would imply that light bending will not significantly enhance the observed flux at low $i$. The second simplifying assumption made here is that the impinging flux is radiated isotropically. This would be valid for the case when the incident radiation is absorbed and re-radiated, but instead, if the radiation is scattered off the surface then the reflected radiation could be anisotropic. However, for Thomson scattering the difference in the differential cross-section between forward/backward scattering and scattering at right angles, is only a factor of 2. Indeed, detailed calculations of the angle dependent reflected spectra shows variations within a factor of a few (Magdziarz & Zdziarski 1995). Thus this effect could at most effect

– 6 –

the calculations by a similar factor and the main conclusions of this work should remain unchanged.

It should be emphasized that the results of this work argues against sub-Eddington thick accretion disk models for ULX. However, ULX could indeed be super-Eddington thick (or thin) accretion disks around stellar mass black holes or they could be accretion disks harboring intermediate size black holes.

K.S acknowledges the Visiting Student Program at IUCAA.

## REFERENCES

Begelman, M. C., 2002, ApJ, 568, L97.

Borozdin, K., Revnivtsev, M., Trudolyubov, S., Shrader, C., & Titarchuk, L., 1999, ApJ, 517, 367.

Colbert, E. J. M. & Mushotzky, R. F., 1999, ApJ, 519, 89.

Colbert, E. & Ptak, A., 2002, ApJS, *in press* (astro-ph:0204002)

Kaaret, P., *et al.*, 2001, MNRAS, 321, L29.

King, A. R., Davies, M. B., Ward, M. J., Fabbiano, G., & Elvis, M., 2001, ApJ, 552, L109.

Kubota, A., *et al.*, 2001, ApJ, 547, L119.

Madau, P., & Rees, M. J., 2001, ApJ, 551, L27.

Magdziarz, P., & Zdziarski, A. A., 1995, MNRAS, 273, 837.

Matsumoto, H., *et al.*, 2001, ApJ, 547, L25.

Portegies Zwart, S. F., & McMillian, S. L. W., 2002, ApJ, *in press* (astro-ph:0201055)

Roberts, T. P. & Warwick, R. S., 2000, MNRAS, 315, 98.

Shrader, C., & Titarchuk, L., 1998, ApJ, 499, L31.

Taniguchi, Y., Shioya, Y., Tsuru, T. G. & Ikeuchi, S., 2000, PASJ, 52, 533.

Zezas, A., & Fabbiano, G., ApJ, *submitted* (astro-ph:0203176)

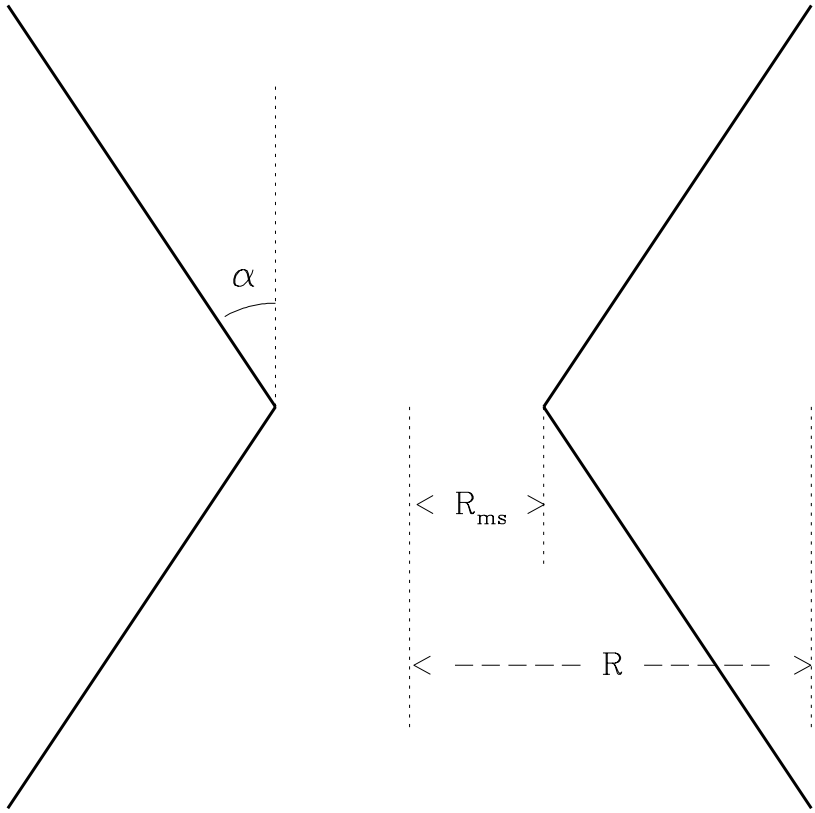

Fig. 1.— Schematic diagram (not to scale) for the geometry of the thick accretion disk. The black hole is in the center of the figure (not shown) and the solid lines mark the outer surface of the thick accretion disk.

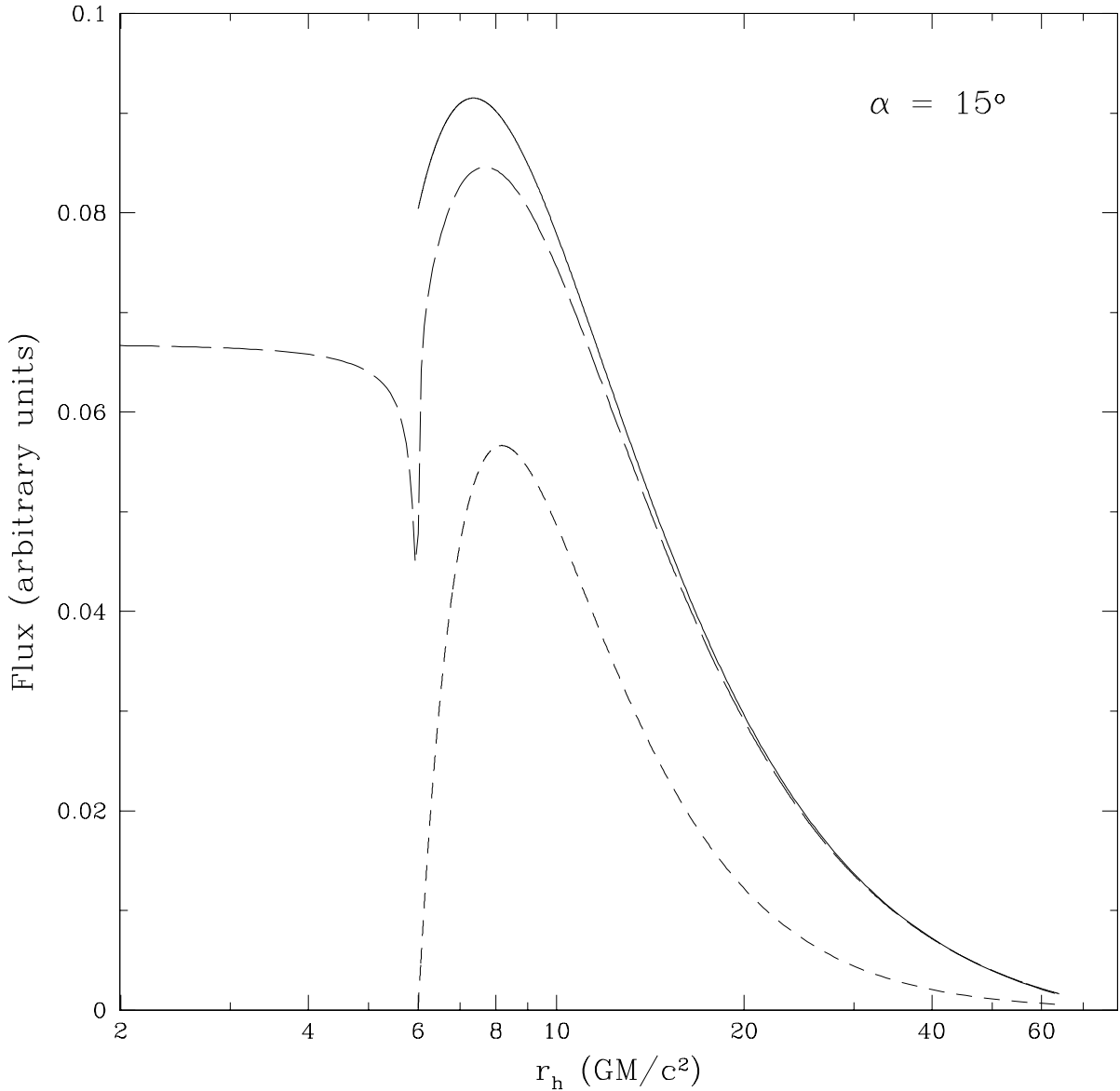

Fig. 2.— The surface flux ($F(r)$) versus the horizontal distance from the black hole ($r_h$) for an opening angle $\alpha = 15^o$. The solid (dashed) curve is for the case when the inner region of the disk ($r_h < r_{ms} = 6GM/c^2$) is optically thin (thick). The dotted curve is the dissipated flux, $F_d$ (eqn. 2) or equivalently the surface flux for a flat disk.

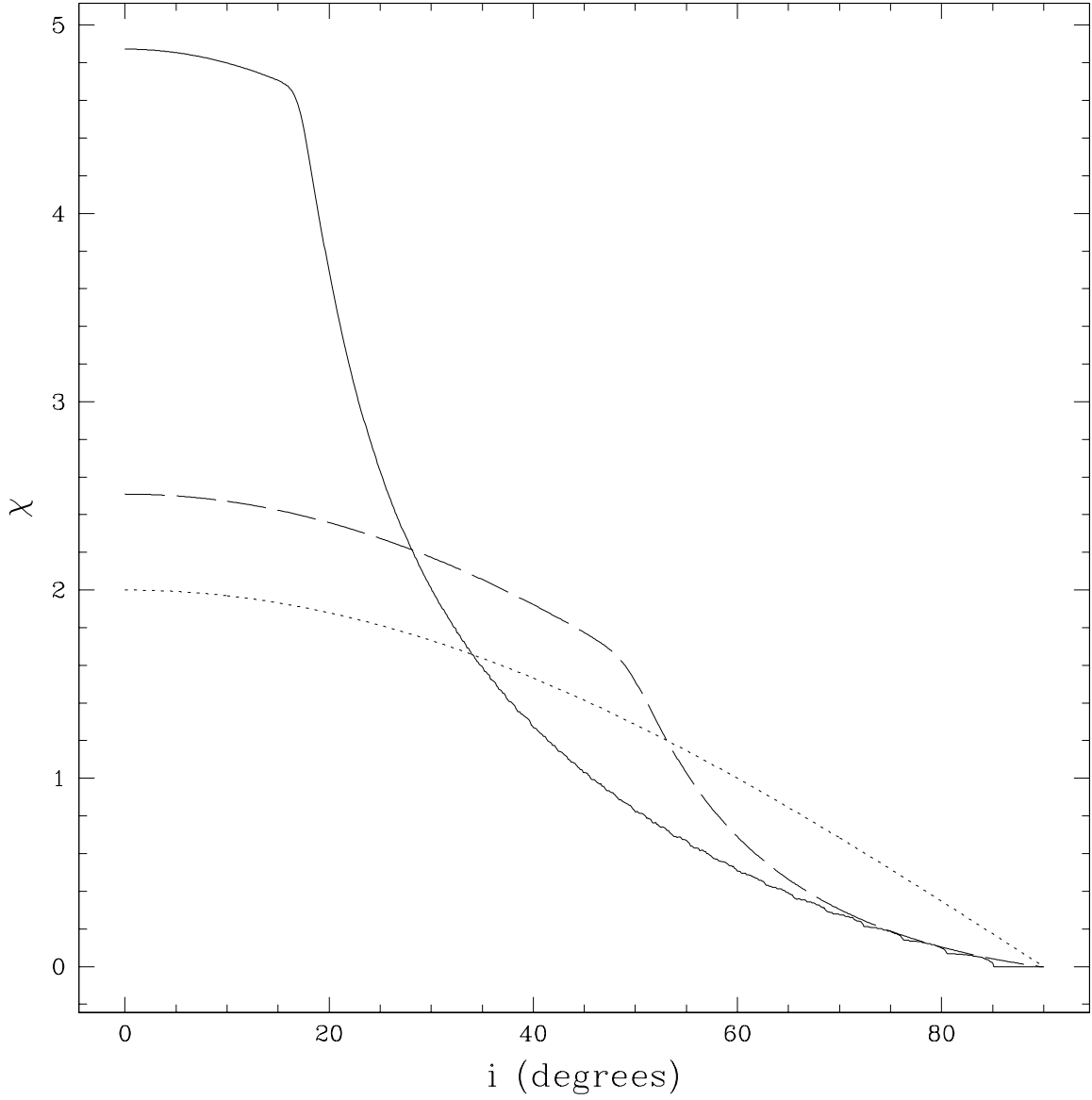

Fig. 3.— The enhancement factor, $\chi$ (which is the ratio of the apparent to real luminosity) as a function of inclination angle, $i$. Solid curve: $\alpha = 15^o$. Dashed curve: $\alpha = 45^o$. Dotted curve is for a flat disk ($\chi = 2\cos i$).